\documentclass[pdflatex,sn-mathphys-num]{sn-jnl}

\usepackage{graphicx}
\usepackage{dcolumn}
\usepackage{epstopdf}

\usepackage[labelfont={bf,sf}, textfont={sf},  justification=raggedright, singlelinecheck=false]{caption}

\usepackage{multirow}%
\usepackage{amsmath,amssymb,amsfonts}%
\usepackage{amsthm}%
\usepackage{mathrsfs}%
\usepackage[title]{appendix}%
\usepackage{xcolor}%
\usepackage{textcomp}%
\usepackage{manyfoot}%
\usepackage{booktabs}%
\usepackage{algorithm}%
\usepackage{algorithmicx}%
\usepackage{algpseudocode}%
\usepackage{listings}%

\usepackage{soul}

\theoremstyle{thmstyleone}%
\theoremstyle{thmstyletwo}%

\theoremstyle{thmstylethree}%

\begin{document}

\title[The mechanistic origin of branching-driven nucleation in abrupt phase transitions]{The mechanistic origin of branching-driven nucleation in abrupt phase transitions}



\author[1,2,3]{\fnm{Leyang} \sur{Xue}}
\author[3,4]{\fnm{Shengling} \sur{Gao}}
\author[5,6]{\fnm{Bnaya} \sur{Gross}}
\author[7]{\fnm{Orr} \sur{Levy}}
\author*[4]{\fnm{Daqing} \sur{Li}}\email{daqingl@buaa.edu.cn}
\author[1,2]{\fnm{Zengru} \sur{Di}}
\author*[8]{\fnm{Lazaros K.} \sur{Gallos}}\email{lgallos@gmail.com}
\author*[3]{\fnm{Shlomo} \sur{Havlin}}\email{havlins@gmail.com}

\affil[1]{\orgdiv{International Academic Center of Complex Systems}, \orgname{Beijing Normal University}, \orgaddress{\city{Zhuhai}, \postcode{519087}, \country{China}}}

\affil[2]{\orgdiv{School of Systems Science}, \orgname{Beijing Normal University}, \orgaddress{\city{Beijing}, \postcode{100875}, \country{China}}}

\affil*[3]{\orgdiv{Department of Physics}, \orgname{Bar-Ilan University}, \orgaddress{\city{Ramat-Gan}, \postcode{52900},  \country{Israel}}}

\affil[4]{\orgdiv{School of Mathematical Sciences}, \orgname{Beihang University}, \orgaddress{\city{Beijing}, \postcode{100191},  \country{China}}}

\affil[5]{\orgdiv{Network Science Institute}, \orgname{Northeastern University}, \orgaddress{\city{Boston}, \postcode{02115}, \state{MA}, \country{USA}}}

\affil[6]{\orgdiv{Department of Physics}, \orgname{Northeastern University}, \orgaddress{\city{Boston}, \postcode{02115}, \state{MA}, \country{USA}}}

\affil[7]{\orgdiv{Faculty of Engineering}, \orgname{Bar-Ilan University}, \orgaddress{\city{Ramat-Gan}, \postcode{52900},  \country{Israel}}}

\affil*[8]{\orgdiv{DIMACS}, \orgname{Rutgers University}, \orgaddress{ \city{Piscataway}, \postcode{08854}, \state{NJ}, \country{USA}}}

\abstract{
Phase transitions are the macroscopic manifestation of microscopic processes that drive a system towards a new state. The detailed evolution of these processes – particularly in abrupt phase transitions -- are currently not fully understood. Here, we introduce a theoretical framework based on internal node dependencies within a single-layer lattice. Crucially, we demonstrate that the fundamental mechanism underlying abrupt transitions is nucleation propagation preceded by a slow cascading process which scales with the range of dependencies. Our findings show that the synergy between these two distinct stages is essential for the occurrence of an abrupt transition. The first stage of a slow cascading mechanism was recently observed experimentally in superconducting layered materials -- where heat acts as the dependency links – for the limit of infinite dependency range. Our model thus generalizes the framework to include finite dependency ranges, revealing previously unobserved mechanisms that could be experimentally verified through controlling the range of thermal diffusion in the material. As a universal mechanism, our model provides a robust method to test nucleation-controlled phase transitions in multiple systems, providing a path to discover and understand microscopic mechanisms in phase transitions. 
}




\maketitle

Phase transitions (PTs) are fundamental for describing how complex systems undergo dramatic changes in their macroscopic 
properties at critical points~\cite{stanley1971phase,domb2000phase,Binder_1987, Dorogovtsev2008, sole2011phase}.
The nature of a phase transition is shaped by the microscopic interactions between elements within the system. An extensive number of interaction types have been studied in the framework of complex networks --- where interactions between nodes can be cooperative~\cite{Cai2015, battiston2021physics}, dependent~\cite{Der2005, buldyrev2010catastrophic, Baxter2015, radicchi2015percolation}, competitive~\cite{achlioptas2009explosive,Nagler2012}, or inhibitory~\cite{sun2023dynamic}. Similarly, the type of the transition is influenced by the range of these interactions~\cite{Gao2024, xue2024nucleation}, which can be either short range, intermdediate range, or long range. 
In a second-order PT, the order parameter changes continuously and exhibits scaling behavior near the critical point.
In contrast, a first-order PT features an abrupt jump at the critical point without any scaling behavior. Recently, a mixed-order PT has received attention, as  it combines an abrupt change, characteristic of first-order transitions, with a scaling behavior near the critical point, similar to second-order transitions~\cite{lee2017universal, mukamel2024mixed}.

Percolation is a well-established theory that yields phase transitions in networks \cite{bollobas2006percolation, cohen2010complex, bunde2012fractals, stauffer2018introduction}. In its simplest form, a fraction of nodes are randomly removed from a network and the behavior of the order parameter is observed. In percolation, the order parameter corresponds to the size of the giant connected component (GCC), which determines the macroscopic state (order) of the system. In traditional percolation studies, this is a static process and the removal of the initial nodes is followed by the immediate removal of finite isolated clusters, yielding a continuous second-order PT as the number of initially removed nodes increases. In contrast, the extensive study of {\it interdependent networks} \cite{buldyrev2010catastrophic} has demonstrated that a percolation framework can yield diverse types of PT phenomena by connecting networks via dependency links. Building on recent observations \cite{bonamassa2023interdependent} of two types of interaction in multi-layered superconductors -- where electrical current acts as the connectivity links in each layer and thermal dissipation creates the dependency links between the layers -- we propose a corresponding theory for percolation in single-layer lattices with two types of interactions where the length of dependency links can be controlled.. This work provides the necessary groundwork for upcoming experiments involving single-layer superconducting systems with controlled range thermal dependency links ~\cite{buldyrev2010catastrophic, bonamassa2023interdependent, artime2024robustness, gao2011robustness}.

Here, we address the question of how {\it internally dependent} and connected nodes in a network cause the collapse of the system due to random failures. Surprisingly, as we show below, this analysis allows us to deeply identify novel microscopic mechanisms of abrupt transitions and particularly to answer the long-standing open problem of what distinguishes between first-order and mixed-order PT, based on spontaneous cascading mechanisms at the microscopic level.

\section*{The dependency model}\label{sec2}
We consider a two-dimensional square lattice of size $N=L \times L$, where each node, located on a lattice site, interacts with other nodes via two distinct types of links: {\it connectivity links},  which form the square-lattice network structure by representing physical or informational connections and determine which nodes remain in the GCC, and {\it dependency interacting links}, which establish a pairwise functional relationship where a node's operation depends on the existence of a specific other node. When a node fails, then its dependent node also fails. This simple additional interaction leads to an interplay between percolation and dependency failures, where the initial removal of nodes can drive a spontaneous cascading failure process and different types of abrupt collapse occur.

\begin{figure}[p!]
    \centering
    \includegraphics[width=\columnwidth]{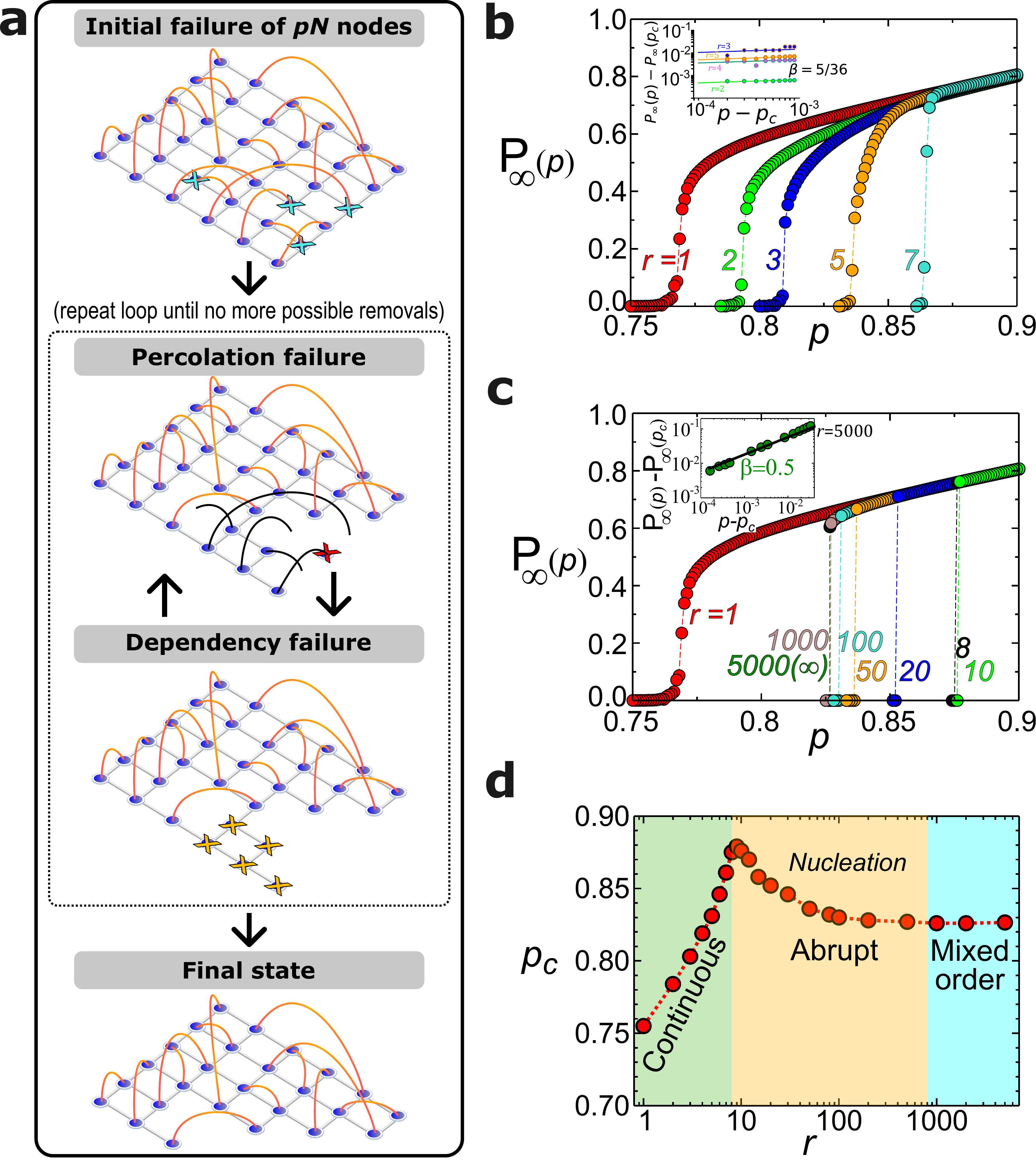}
    \caption{ {\bf Distinct phase transitions in the internally dependent lattice.}
    {\bf a,} Schematic of the model, where dependency links are shown in orange. Initially, we remove a fraction of nodes, marked with cyan crosses at the top panel. Following this removal, we repeatedly iterate two stages. At the percolation failure stage, we remove nodes that become disconnected from the GCC, marked with a red cross in the second panel. This is followed by the dependency failure stage, where we remove nodes that depend on the lost nodes (yellow crosses). The process continues by alternating the percolation and dependency failure stages until a steady state is reached.
    {\bf b,} The final size (after cascading stops) of the GCC, $P_{\infty}(p)$, as a function of the initial non-removed probability, $p$, for small values of $r\leq 7$, as indicated in the plot. All the transitions are continuous and the inset shows that the power-law exponent, $\beta$, in the vicinity of the transition has the same value of $\beta=5/36$ as in regular two-dimension percolation. {\bf c,} The final size of the GCC, $P_{\infty}(p)$, as a function of the initial non-removed probability, $p$, for larger values of $r\geq 8$, as indicated in the plot. All the transitions are abrupt. For larger $r$ values, of the order of the system size, the curvature close to the critical point indicates a mixed-order transition with critical exponents. As shown in the inset, the exponent for the mixed-order transition at $r=5000$ has a value $\beta=1/2$, see Eq.~\ref{Eq:beta}.
    {\bf d,} The phase diagram and the critical threshold $p_c$ versus the dependency length $r$.
    All results in this Figure are for networks with size $N=L \times L=5000\times 5000$.
    }
    \label{fig:1}
\end{figure}

To account for the spatial range of dependency interactions, we introduce a tunable parameter $r$, which defines the dependency interaction range, allowing a node to depend on another single node at a distance smaller or equal to $r$. We treat the special case of $r=L$ as the equivalent of `infinite-range' dependency, $r=\infty$, where dependency links are randomly distributed throughout the entire network, independently of distance. An example of this model is illustrated in Fig.~\ref{fig:1}a.

The process begins with a random removal of a fraction $1-p$ of nodes from the network. This leads to removing nodes that become disconnected from the GCC. In traditional percolation, this would conclude the removal process and we would simply need to calculate the GCC size, $P_\infty$. However, the dependency links now introduce an additional failure mechanism, which gives rise to a dynamical cascading process. After the initial removal of nodes, either because of initial failure or because of GCC disconnection, all nodes in the network depending on the removed nodes will also fail, regardless of their connectivity within the network. This would lead to removing new disconnected clusters that were connected to the GCC via nodes that lost their dependency links, and so on. These two alternating removal mechanisms form a cascading process, which continues until the system collapses or stabilizes and a finite GCC is formed.

Figures~\ref{fig:1}b-c show that the dependency length $r$ controls the phase transition behavior, shifting it from one PT type to another. For small values of $r$, the system undergoes a continuous phase transition, so that for $p$ above $p_c$:
\begin{equation}
P_\infty(p)-P_\infty(p_c) \sim (p - p_c)^\beta \,.
\label{Eq:beta}
\end{equation}
The above equation defines the scaling exponent $\beta$. For $r<8$ we find that $P_\infty(p_c)=0$ and the value of the exponent is $\beta=5/36$, same as traditional percolation \cite{bunde2012fractals}, as shown in the inset of Fig.~\ref{fig:1}b. As we increase $r$ up to $r=7$, the value of $p_c$ increases (Fig.~\ref{fig:1}d), with the PT remaining continuous. However, when $r$ reaches a critical threshold $r_c=8$, $P_{\infty}(p_c)$ exhibits a sudden jump at the critical point $p_c$, signaling an abrupt phase transition. This abrupt jump remains for all values of $r\geq r_c$. Note that for $r_c<r\ll L$, the abrupt transition does not involve a curvature near $p_c$  (Fig.~\ref{fig:1}c) and therefore there is no critical exponent near $p_c$. In contrast, as $r\to L$, a noticeable curvature appears near and above $p_c$ in $P_\infty$ vs $p$ (see Fig.~\ref{fig:1}c, $r$=1000, 5000), indicating the presence of critical behavior with an exponent close to the theoretical value of $\beta=1/2$ (shown in the inset). This result points to the existence of a mixed-order transition \cite{lee2017universal,Gross2022,mukamel2024mixed}.

Why do longer dependency links change the order of the transition from first-order to mixed-order?
As seen in Figs.~\ref{fig:1}b and \ref{fig:1}c, the GCC size at large $p$ values, $p>p_c$, collapses into a single universal curve independently of $r$ (from $r=1$ to $r=L$).
While this convergence may suggest a similar structure for any $r$ value, the GCC size is a macroscopic metric that only quantifies the average damage in the system. The divergence of these curves near their respective critical points $p_c(r)$ reveals that the internal structure varies significantly at the microscopic level, as we will show below. Specifically, the range of dependency links alters the microscopic failure propagation mechanisms: short-range links favor local and continuously growing nucleation-type damage, while long-range links lead to abrupt cascades across the entire system. The varying microscopic dynamics is the main factor that drives the order and the nature of the phase transition.

\section*{Results}\label{sec3}

\subsection*{Transition Mechanisms}

The collapse process at the critical threshold, $p_c$, is strongly dependent on $r$. Figure~\ref{fig:2} shows the time evolution (cascading steps) of $P_\infty$ at $p_c$ for several $r$ values, where one time step represents a single percolation and dependency failure cycle. We identify four primary regimes (described in detail in the Supplementary Information, SI): (i) For $r < r_c (=8)$, the breakdown process resembles a standard second-order percolation transition.
(ii) When $r$ is less but close to the critical value, i.e. for $r \lesssim  r_c$ such as $r=7$, the system collapses continuously while $p_c$ increases (with $r$), as multiple holes grow simultaneously across the system. (iii) For values equal or larger than the critical value $r_c(=8)$, but smaller than the system size, e.g. $r=50$, the system collapses in two stages: first, a cascading process acts at a microscopic level (order $r$) by removing only a microscopic number of nodes per step around one or few locations, demonstrated by low-density small regimes in Fig.~\ref{fig:2}a and a plateau in the GCC size up to about $t=15$ (Fig.~\ref{fig:2}b). The second stage, for $t>15$, is characterized by an accelerating collapse as an increasingly larger number of nodes are removed, demonstrated by the propagation of the hole, that is a nucleation process shown in Fig.~\ref{fig:2}a and the sharp decrease of $P_\infty(t)$ in Fig.~\ref{fig:2}b. The duration of the plateau in the first stage increases monotonically with increasing $r$, as $r^{2/3}$, see Eq.~\ref{Eq:27} and Fig.~\ref{fig:3}a-b. (iv) As $r$ approaches the system size, i.e. $r\to L$, the plateau is the only stage that occurs, and the system collapses abruptly in a couple of steps.

\subsection*{Nucleation process}
For $r_c\leq r\ll L$, the spontaneous {\it branching process} during the breakdown at $p_c$  starts with a plateau stage, as shown for $r=50$ and $r=500$ in Fig.~\ref{fig:2}b. A very small fraction of nodes are removed per step, on the order of $10^{-4}$ to $10^{-5}$, as can be seen in Fig.~\ref{fig:2}c. This stage occurs for only a few tens of steps, which clearly suggests that there are no system-wide removal avalanches. Instead, damage remains localized within a radius $r$ around the initially failed nodes. Consequently, the density throughout the system remains relatively high, as depicted in (A) and (B) of Fig.~\ref{fig:2}a, and the large-scale connectivity is preserved until the end of this plateau stage. Note, however, that early in this plateau, e.g. at $t=5$ (Fig.~\ref{fig:2}a, (A)), spontaneous cascade causes damage to only one or few local areas of order $r$ that create a {\it nucleation} zone with significantly lower local density. When $p>p_c$, but close to $p_c$, the density within these potential nucleation areas is low compared to other areas, but still high enough to end the cascading process, so that nucleation propagation will be short and collapse does not occur. However, at the critical point $p=p_c$ the density surrounding the nucleus (usually a single one dominates) drops sufficiently and beomes a `hole', and initiates a {\it nucleation propagation}, as shown in panel (B) of Fig.~\ref{fig:2}a. The initial small `hole', with size of order $r$, expands outwards as nodes in the perimeter fail due to the dependency length $r$ (panels (C) and (D) in Fig.~\ref{fig:2}a).  In the absence of nucleation propagation ($p>p_c$), the branching process stops with a finite-size GCC, whose structure is far from a critical state. In contrast, at $p=p_c$ the initial nucleation event can propagate indefinitely, inducing a system-wide collapse. This triggers the first-order transition and results in a significantly higher critical threshold, $p_c$, compared to classic percolation with no dependencies.

As $r$ increases, the overall collapse mechanism remains the same, but the system nucleation conditions require a larger area of radius of order $r$ to become diluted and form a hole. Consequently, a larger initial damage, 1-$p$, is required to trigger system collapse, resulting in a lower critical point, $p_c$, as indeed seen in Figs.~\ref{fig:1}c-d. This picture can also explain why for larger $r$, nucleation is delayed (longer plateau, see Fig.~\ref{fig:2}b), emerging after significant damage spreading, which leads to a scaling law of steadily increasing duration for the plateau stage, see Fig.~\ref{fig:3}a-b and Eq.~\ref{Eq:27}. Note, that the {\it nucleation} transition is analogous to gas-liquid transitions~\cite{glaser1952some}, although the underlying mechanisms differ, i.e., while surface tension is believed to be the driving force in gas-liquid transitions~\cite{glaser1952some}, here the nucleation is driven by a failure front determined by the dependency length scale.

The nucleation propagation mechanism, which drives the abrupt transition, can be supported  by tracking the failed giant connected component, $P^f_\infty (t)$ (failed-GCC; see Methods). As shown in Figure~\ref{fig:2}d, failed nodes are initially distributed randomly with no significant failed clusters. However, this stage is followed by the emergence of a large failed-GCC that continues to grow in size until it spans the entire system. The only explanation of this behavior is that, after a certain delayed time, $\tau_b$,  which increases with $r$ as Eq.~\ref{Eq:27},  failures are highly correlated locally in space and form a tight expanding hole. Note that the duration of this propagation process decreases with increasing $r$, for the same $L$, vanishing almost entirely as $r \sim L$, a regime which is discussed in the nucleation duration section below.

\subsection*{Random cascading}
As $r$ approaches the system size ($r\lesssim L$), the failure mechanism during the plateau stage remains the same as described above, but now the emergent nucleus covers the {\it whole} system.
The removal of a node, in this case, can induce damage anywhere randomly in the system, so that a nucleation zone with characteristic length $r\sim L$ emerges and  spans the entire lattice system. Naturally, in this case, any emergent nucleus cannot propagate, since its existence immediately triggers a total system collapse. This condition requires a larger initial damage (that is, a smaller $p_c$) compared to the smaller $r$ cases, because of the size of the nucleus.
As $p$ approaches $p_c$ from above, cascading failures lead to progressively greater accumulated damage almost uniformly throughout the system. Since $r$ is the largest, this damage leads to a longer plateau stage. Consequently, as the initial damage increases with decreasing $p$, the size of GCC for $p>p_c$ decreases and curves downwards in contrast to $r\ll L$ cases.
These long-range cascades are represented by the critical curvature of $P_\infty$ seen near $p_c$ for $r\sim L$ with a scaling exponent $\beta=1/2$, see Fig.~\ref{fig:1}c. This explains the scaling features analogous to a second-order transition for $p\gtrsim p_c$ (see inset of Fig.~\ref{fig:1}c and Eq.~(\ref{Eq:beta})) and the abrupt collapse at $p=p_c$. In other words, for $r=L$ the emergent `hole' already covers the entire system, yielding the curvature, critical scaling, and thus demonstrating and recovering the mechanisms behind the {\it mixed-order transition}. In contrast, at smaller $r$ values there is no curvature in $P_\infty(p)$, because the creation of the nucleus capable of propagating and collapsing the system -- a hallmark of a typical {\it first-order transition} -- only requires damage within a small localized area of order $r\ll L$, whose size is negligible compared to system size and can not affect the giant component $P_\infty$ (see Fig.~\ref{fig:1}c for $r<100$).

\subsection*{Plateau duration}
Analysis of the duration of the plateau and nucleation phases reveals novel scaling laws. Extensive  simulations show that at $p=p_c$ the mean plateau duration $\langle \tau_b \rangle$ (see Fig.~\ref{fig:2}b) and the corresponding standard deviations, scale with the dependency length $r$ as: 
\begin{equation}\label{Eq:27}
\langle \tau_b \rangle \sim r^{2/3}, \quad \sigma(\tau_b) \sim r^{2/3}.
\end{equation}
The distribution of $\tau_b$  collapses when using $P(\tau_b)r^{2/3} \sim F(\tau_b/r^{2/3})$, where $F$ follows a skewed Gaussian profile~(Fig.~\ref{fig:3}a), supporting the validity of the scaling relations in Eq.~(\ref{Eq:27}), which are also directly supported by Fig.~\ref{fig:3}b.

For the limiting case of $r=L$, the above scaling relationships have an exponent 1/3 as a function of $N=L^2$, i.e. $\langle \tau_b \rangle \sim N^{1/3}$. That is, we find in the present study that Eq.~(\ref{Eq:27}) is {\it general} for any $r$ value and for $r=L$ it yields the specific case  $\langle \tau_b \rangle \sim L^{2/3} \sim N^{1/3}$ \cite{gross2025random}.
Furthermore, the $N^{1/3}$ scaling is consistent with that observed across various interdependent networks, including abstract interdependent networks~\cite{zhou2014simultaneous}, experimentally for interdependent superconducting networks~\cite{gross2025random} and interdependent ferromagnetic networks~\cite{gross2024microscopic}.
Thus, our current study suggests that the common scaling exponent points towards a universal underlying mechanism in systems with {\it finite} and {\it infinite} dependency interactions, independent of the network structure or the physical process.

\subsection*{Nucleation duration}
\begin{figure}[p!]
    \centering
    \includegraphics[width=\columnwidth]{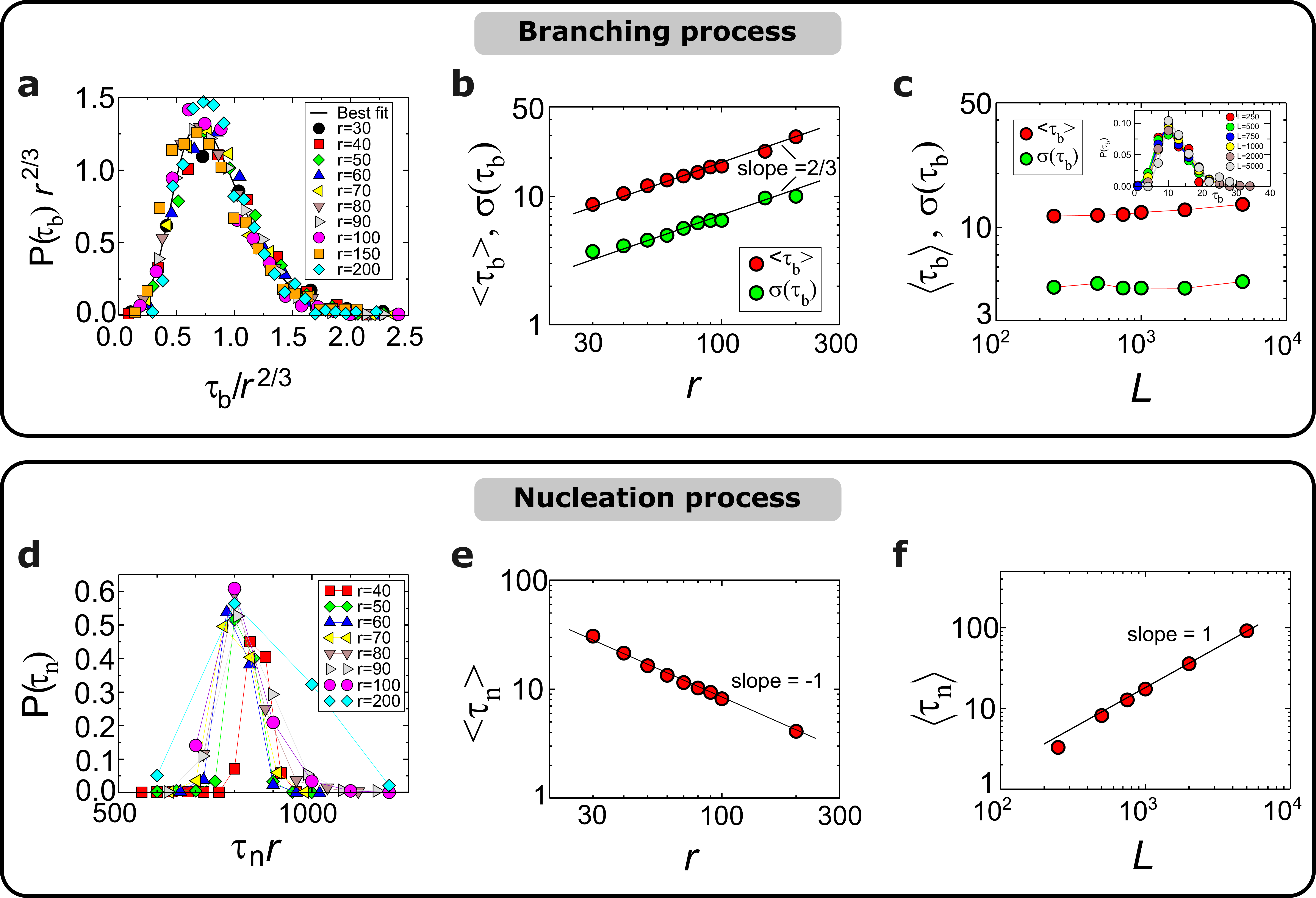}
    \caption{
    {\bf Scaling behavior of abrupt cascade failures during the branching process (top row) and during the nucleation process (bottom row).}
     {\bf a,} Scaling collapse of the distribution $P(\tau_b)$ of the branching plateau duration, $\tau_b$, at $p_c$. {\bf b,} Scaling of the mean and standard deviation for the branching duration $\tau_b$ with $r$. Note that the scaling with $r$ is the generalization of the scaling with $L$ for $r=L$, found experimentally and theoretically see \cite{bonamassa2023interdependent,zhou2014simultaneous}. {\bf c,} The dependence of  $\langle \tau_b \rangle$ and $\sigma(\tau_b)$ on $L$, for fixed $r$, e.g. $r=50$, shows independence on $L$. This shows that the emergence of nucleation does not depend on the size of the system. Inset: The distribution of $\tau_b$ remains the same across different system sizes. {\bf d,} Scaling collapse of the nucleation duration distribution $P(\tau_n)$. {\bf e,} Scaling of the nucleation duration, $\tau_n$, with $r$. {\bf f,}  Scaling of $\langle \tau_n \rangle$ with $L$ for $r=50$.
    Each point represents an average over 1000 runs. The system size is $L=1000$, except for panels (c) and (f).
    }
    \label{fig:3}
\end{figure}

Scaling relations extend also to the duration of the nucleation propagation stage, with a different scaling exponent. As shown in Fig.~\ref{fig:3}d, the distributions of the nucleation duration for different sizes collapse as $P(\tau_n)\sim F(\tau_n/r^{-1})$, with a scaling exponent -1. That is, the
mean nucleation duration, $\langle \tau_n \rangle$, for a given $L$ scales as $\langle \tau_n \rangle \sim r^{-1}$
(Fig.~\ref{fig:3}e).
This value can be theoretically predicted, based on the observation above that, in nucleation, the hole propagation speed, $v$, is proportional to the dependency length~($v\sim r$).
Thus, for a system of fixed length $L$, the nucleation duration $\tau_n$ scales as $\tau_n \sim L/v \sim L/r$, leading to the -1 exponent.

One of the key distinctions between the plateau stage and the nucleation propagation stage is their dependence on the linear size $L$ of the lattice. For instance, the values of $\langle \tau_b \rangle$ and $\sigma(\tau_b)$ remain constant as we vary $L$ (see e.g., $r=50$, Fig.~\ref{fig:3}c). This strongly supports our observation in Fig.~\ref{fig:2}a that the branching process, for finite $r$, operates at a {\it microscopic} level, mainly at a random but specific location, since the emergence of a low-density regime of size $r$ is not influenced by the system size.
In contrast, the nucleation propagation stage is \textit{macroscopic}, since the hole spreads through the entire system and the GCC size decreases macroscopically, as seen for e.g. $r=50$ in Fig.~\ref{fig:2}b. As a result, the duration of nucleation propagation now scales linearly with the system size, as indeed seen in Fig.~\ref{fig:3}f.

The above findings can better clarify and resolve the open general question about what distinguishes pure first-order transitions from mixed-order transitions. The  curvature and the critical behavior near $p_c$ for mixed-order is due to cascading that occurs for $r=L$ over the whole system, thus effecting the curvature in the giant component and make the transition critical (the whole system plays a role). In marked contrast, for $r\ll L$ the cascading occurs at a small area at a random location (of lowest density and of zero fraction of the system) and therefore the giant component is not influenced and critical exponents do not appear. See e.g. in Fig.~\ref{fig:1}c the distinction between $r=1000$, 5000 and $r<100$, where the curvature near $p_c$ disappears and therefore critical exponents do not exist.

\subsection*{Branching factor}
We showed above that both finite (first-order) and infinite range (mixed-order) dependency systems go through a slow branching process (which for finite $r$ is followed by the nucleation propagation stage), but how different are these processes at the microscopic level?  To gain insight into the underlying mechanisms, we calculate the branching factor, $\eta$, defined, during the plateau cascade, as the ratio between newly failed nodes and the number of failed nodes in the preceding step (see Supplementary Information).

Figure~\ref{fig:4}a shows that for $r=L$ the average branching factor is $\langle \eta \rangle \approx 1$ at $p_c$. This indicates a critical branching process where failures propagate at a sustained, critical rate, placing the system on the brink of collapse. In contrast, below $p_c$~($\Delta p<0$), $\langle \eta \rangle$ exceeds 1, signifying an accelerating cascade (thus, a shorter plateau), while for $\Delta p > 0$,  $\langle \eta \rangle$ becomes below 1, leading to a gradual slowing down of cascading failures, cascades are finite and the giant component remains stable. A larger $|\Delta p|$ yields a larger deviation from 1 (see Fig.~\ref{fig:4}a). These findings support further the conclusion that for infinite $r$, the branching process is the only driver of system collapse.
The system at $p_c$ just at the verge of collapsing, reaches a state where any {\it finite} fraction of failed nodes causes an infinite cascade with branching factor above 1 that yields the collapse of the system.

Surprisingly, the values of $\langle \eta \rangle$ at $p_c$ for finite $r$, are significantly below 1 and approach 1 as $r\rightarrow\infty$ (Fig.~\ref{fig:4}b), showing that the system would not collapse if the branching process was the only mechanism at play. However, as shown above, this branching process leads, somewhere in the system, to a locally denser failure regime of scale $r$ which is sufficient to trigger the second mechanism, i.e. a macroscopic nucleation propagation process. Interestingly, the branching factor remains $\langle \eta \rangle <1$, even when $p<p_c$, so that nucleation is still the main mechanism that leads to the eventual collapse.

The role of $r$ can also be understood by observing how far the damage spreads from the initial failure. Fig.~\ref{fig:4}c shows the distance, $d$, of failed nodes from the center of the nucleus at a given step during the branching process, showing that most of failures occur near the nucleation center. This failure distance increases linearly with $r$, suggesting, as expected, that longer dependency links lead to more spread damage. Conversely, at small $r$, failures take longer to propagate far, resulting in spatially limited collapse patterns.

\section*{Discussion}

We presented a novel framework that fundamentally enlightens our understanding of the origin of different types of abrupt phase transitions. Unlike traditional percolation, the introduction of internal dependencies within a single-layer network system creates a dynamic cascading process, governed by the dependency range. This framework represents a testbed that can be tuned to yield, analyze, and understand the mechanisms behind all three types of PT, i.e. continuous, abrupt, and mixed-order. Our model is realistic and timely since recent work \cite{bonamassa2023interdependent} studies the behavior of materials with two such types of links, where the dependency links are found to be physical and mediated by heat propagation, as happens in superconducting layered materials \cite{bonamassa2023interdependent} or via light in laser systems \cite{wang2025spontaneouscascadingmechanismcritical}. Thus, the present study  can inspire and guide novel experiments for observing for the first time nucleation in single-layer materials due to limited range of heat dependency that will enable efficient control of their PT behavior.

With this goal in mind, our study reveals that microscopic failure processes are the origin of macroscopic PT characteristics, and clarifies distinctions between first-order and mixed-order PTs, in general, by their microscopic origins. First-order PTs, stemming from finite dependencies, involves a randomly localized damage at a finite scale (i.e. a zero fraction of the system) so that the system remains far from a critical state when the spontaneous nucleation event propagates system-wide. This leads to sudden global failure without curvature or critical exponents of the giant component (see e.g. Fig.~\ref{fig:1}c for $r=50$). In contrast, when $r$ is of the order of $L$, mixed-order PTs arise from long-range dependencies that induce progressive, system-wide weakening above and near $p_c$. This widespread damage is indicative of a mixed-order transition and explains its main features, i.e., the observed curvature near and above $p_c$ and the critical exponents (e.g., Fig.~\ref{fig:1}c for $r=1000$ and 5000, with $\beta=1/2$ in Eq.~(\ref{Eq:beta})). The critical curvature near $p_c$ of the GCC is interrupted by an abrupt collapse at $p_c$, which is the result of an extended cascading process with $\eta=1$, and ultimately leads to the mixed-order PT.

Our findings demonstrate that a realistic extension to existing percolation models, i.e. the addition of dependency interactions, yields surprisingly rich phenomena and complex emergent behaviors that traditional models miss and can be realized in future experiments. Our model can serve as a generic template for multiple dynamic processes in systems with inter-dependent interactions. For example, as can be seen in Fig.~\ref{fig:S2}, we ideintified four different mechanisms behind the system collapse, which can be controlled by the dependency range. In real materials, this would indicate the different pathways behind a phase transition and how the system evolves before reaching the critical point. At the same time, the microscopic structure of the system can be very different at criticality, as seen in Fig.~\ref{fig:S2}, even though the macroscopic measures are quite similar. Our theoretical findings offer a deep and new understanding of the mechanisms behind PTs, showing that spontaneous microscopic processes yield macroscopic changes in ways that could be experimentally observed in diverse natural and technological contexts.

\section*{Methods}

{\bf Simulation setup.} We generate a standard $N=L \times L$ two-dimensional square lattice, where $L$ is between 1000 and 5000. Each node is connected to its four nearest neighbors and we employ periodic boundary conditions. Furthermore, every node depends on exactly one other node in the lattice via a bidirectional dependency link. To assign these dependency links for a given $r$ value, we tested two methods which produced nearly identical results. In the first method, for each node we randomly select a dependency node from all nodes at Manhattan (also called taxicab) distance exactly equal to $r$. In the second method, we select a node uniformly from all nodes within a Manhattan distance $r$ or smaller. In either case, if a node has no  available targets in its range, because all potential targets have been assigned dependencies in previous steps, then it remains without dependency and is removed at the first stage of the failure process. The fraction of these unassigned nodes is very small and diminishes with $r$. When $r=L$, the dependency links are always chosen at random from the entire system and there are no unassigned nodes.

{\bf The failure process.} A node is considered active only if it meets two conditions: a) it is connected to the GCC, and b) its assigned dependent node is also active. The failure process is initiated by randomly removing a fraction $1-p$ of the lattice nodes. This triggers a cascade of failures that continues until the network either achieves a final stable state, where no more nodes fail, or it collapses completely. The cascade proceeds in discrete time steps, $t$, where each step is a two-part cycle. a) Percolation failure. In the first half of the step, we identify the GCC via a standard burning algorithm. All nodes not connected to the GCC, i.e. those in the smaller clusters, are removed. b) In the second half of the step, we identify and remove all remaining nodes whose dependency nodes have failed earlier. This process is iterative. The dependency failures in this step can fragment the GCC and create more isolated clusters so that in the next step these smaller clusters are removed, followed by more dependency removals, and so on.

At each time step we calculate the following quantities: the size of the giant connected component of operational nodes, $P_\infty(t)$, the number of failed nodes at this time step, $S(t)$, and the size of the failed giant connected component, or failed-GCC, $P^f_\infty(t)$. For the calculation of the GCC and the failed-GCC, we only take into account the interaction links, which means that the dependency links do not contribute to the connectedness of the network. For the GCC, we consider all nodes that are operational at time $t$, and the value of $P_\infty(t)$ is the fraction of nodes currently in the GCC relative to the initial network size. For the failed-GCC, we take into account all nodes that have failed until time step $t$ and perform a standard cluster size analysis. The largest cluster of currently failed nodes determines the value of $P^f_\infty(t)$, as the ratio of nodes in this cluster relative to the initial network size. If the removed nodes are scattered randomly within the lattice and only small failure clusters exist, the value of $P^f_\infty(t)$ is close to 0 and it starts increasing when the failed nodes form a large connected cluster, which happens e.g. when a nucleus is formed or the failure small clusters percolate through the lattice. 

{\bf Calculation of the branching factor.}
To measure the branching factor, $\eta$, we start with the stable GCC of the system at the end of the failure process. For $p > p_c$, the GCC includes a non-zero fraction of nodes, and we simply remove a randomly chosen single node from it. We then measure the resulting additional percolation failures occurring in a single step. This process is repeated across multiple independent simulations to compute the average value of $\langle \eta \rangle$ for $\Delta p = p-p_c$ ($>0$).

For $p \leq p_c$, we cannot use the exact same method, because the GCC vanishes. In this case, we bring the system very close to criticality by initially removing a fraction of nodes $1-(p_c+\epsilon)$, where $\epsilon$ is close to 0. We then follow the failure evolution which will bring the system just above the critical point. We then remove an additional fraction of $\Delta p=p-p_c-\epsilon$ nodes from the GCC at criticality, and perform one percolation step where we measure the number of nodes failing at this step. In this case, $\eta$ is calculated as the ratio of additional next-step failures relative to the number of removed nodes, $N \Delta p$. The average value of $\langle \eta \rangle$ is again computed over multiple independent realizations of the system.

\bmhead{Acknowledgements}
This work was supported by the following funding sources: Israel Science Foundation, Grant No.~201/25~(S.H.); Binational Israel-China Science Foundation, Grant No.~3132/19~(S.H.); NSF-BSF, Grant No.~2019740~(S.H.); Israel Ministry of Innovation, Science \& Technology, Grant No.~01017980~(S.H.); Science Minister-Smart Mobility, Grant No.~1001706769~(S.H.); EU H2020 Project OMINO, Grant No.~101086321~(S.H.); The Israeli VATAT project on power-grid~(S.H.); The German-Israel DIP agency (SH);
Key Program of the National Natural Science Foundation of China, Grant No.~71731002~(Z.D.); China Scholarship Council Program~(S.G., L.X.); Israeli Sandwich Scholarship~(L.X.); and computational resources supported by the Interdisciplinary Intelligence SuperComputer Center of Beijing Normal University Zhuhai~(L.X.).


\bibliography{ref}

\clearpage

\begin{figure}[p!]
    \centering
    \includegraphics[width=0.9\columnwidth]{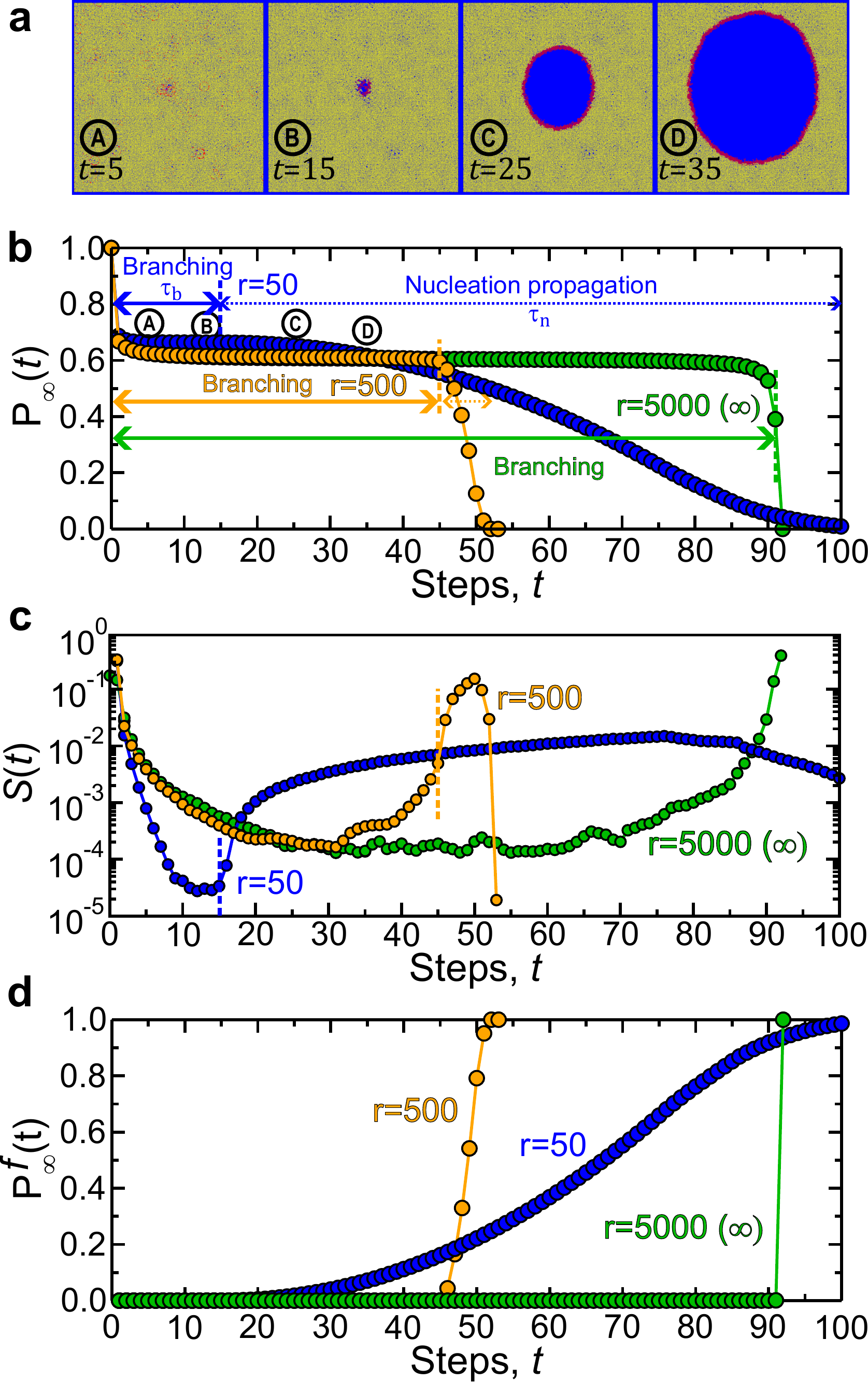}
    \caption{ {\bf The time evolution of the spontaneous failure process shows how localized branching can initiate nucleation propagation.}
    {\bf a,} Snapshots of the largest cluster (yellow), failed nodes in previous steps (blue), and failed nodes at the current step (red),  for $r=50$ at four different times during the cascade at $p_c$, as shown in panel (b). The system size is $L=5000$, but for clarity we zoom in a window of $L=2000$. 
    {\bf b,} The size of the GCC, $P_\infty(t)$, as a function of time (iterations) for $r=50$, $r=500$, $r=5000$ at the critical point $p_c$ during the spontaneous cascade which leads to the abrupt collapse. The dashed vertical lines indicate the point where the plateau branching process ends and the nucleation propagation process starts (there is no nucleation propagation at $r=5000$). Notice that the duration of branching, $\tau_b$, increases with $r$ (Eq.~\ref{Eq:27}), while the duration of nucleation, $\tau_n$, decreases accordingly.
    {\bf c,} Fraction of failing nodes at each time step for the same $r$ values as in panel (b). {\bf d,} Size of the failed-GCC, $P^f_\infty(t)$, as a function of time. The failed-GCC starts with zero size, until a nucleus is formed and steadily propagates radially through the system at different rates.
    The system size is $L = 5000$. Similar plots for smaller $r$ values are presented in the Supplementary Information, Figs. S1 and S2.
    }
    \label{fig:2}
\end{figure}

\clearpage

\clearpage

\begin{figure}[p!]
    \centering
    \includegraphics[width=\columnwidth]{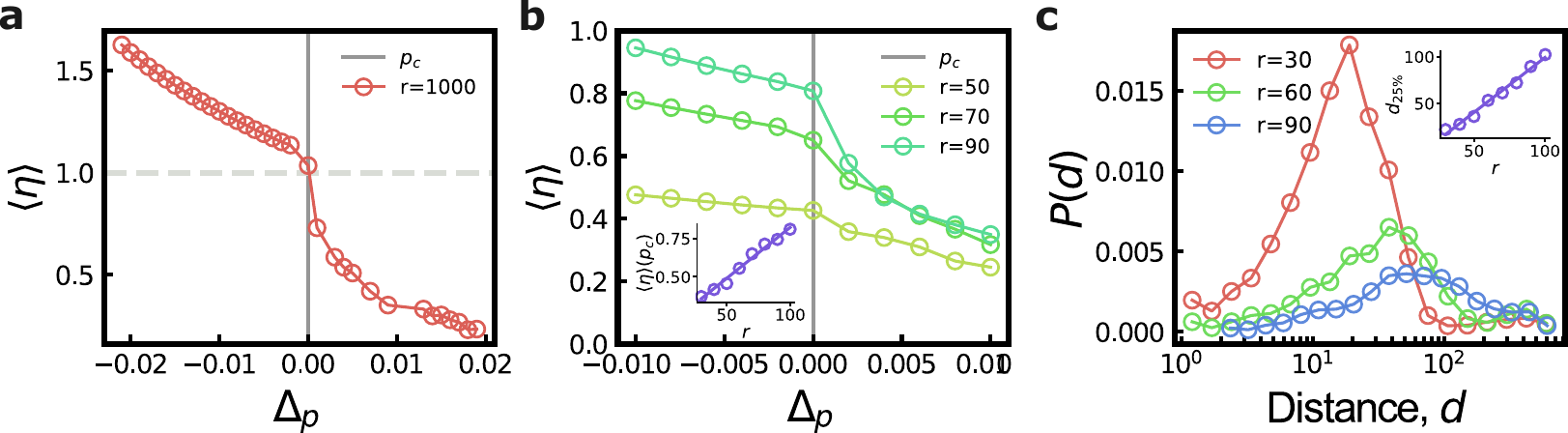}
    \caption{
    {\bf The branching factor controls the system evolution near criticality.} Average branching factor $\langle \eta \rangle$ as a function of $\Delta p= p-p_c$, for {\bf a,} $r=L$, and 
    {\bf b,} for finite-range dependency, $r = 50, 70, 90$. 
    The inset shows that $\langle \eta \rangle(p_c)$ increases linearly with $r$. Note that in contrast to $r=L$, where $\eta=1$ at criticality, for finite $r$, $\eta$ is surprisingly below 1 and approaches 1 for  $r=L$.
    {\bf c,} The distribution of distance, $P(d)$, between failed nodes and the nucleus center at a given step during the branching process, for different $r$, showing that most of failures occur near the microscopic nucleation center. Inset: The 25th percentile value of the distribution of distances between failed nodes and the nucleus center, measured at steps during the branching process at criticality showing linearity with $r$.
    The points represent the average of 10,000 independent runs for a system size of $L=1000$.
    }
    \label{fig:4}
\end{figure}

\clearpage

\setcounter{page}{1}
\setcounter{figure}{0} 
\renewcommand{\thefigure}{S\arabic{figure}}

\begin{center}
    {\Large\bf Supplementary Information}
\end{center}

\subsection*{Time evolution of failures}

As described in the main text, the dependency range, $r$, controls the locality of the failure process and, in turn, the time evolution, the final state, and the order of the phase transition. We identified four distinct regimes of r, as follows:
\begin{itemize}
    \item[(I)] $r<r_c$. In this regime, the phase transition is continuous and all the main features of the PT, including the critical exponents and the fractal character of the GCC structure, are similar to those of a typical percolation PT.
    \item[(II)] $r\lesssim r_c$. Just before the critical dependency range of $r_c=8$, e.g., at $r=7$, the percolation threshold increases significantly, i.e. the system reaches criticality with significantly lower initial damage compared to smaller $r$. The PT remains continuous.
    \item[(III)] $r_c \leq r \ll L$. When $r$ is equal or larger than $r_c=8$, but significantly smaller than the system size, the PT becomes abrupt and exhibits all the features of a first order PT.
    \item[(IV)] $r\sim L$. As the dependency length approaches the system size, the transition remains abrupt, but the order parameter, i.e. the GCC size, decays continuously just before the abrupt jump. This is the hallmark of a mixed-order transition.   
\end{itemize}

\begin{figure}[hbt]
    \centering
    \includegraphics[width=.7\columnwidth]{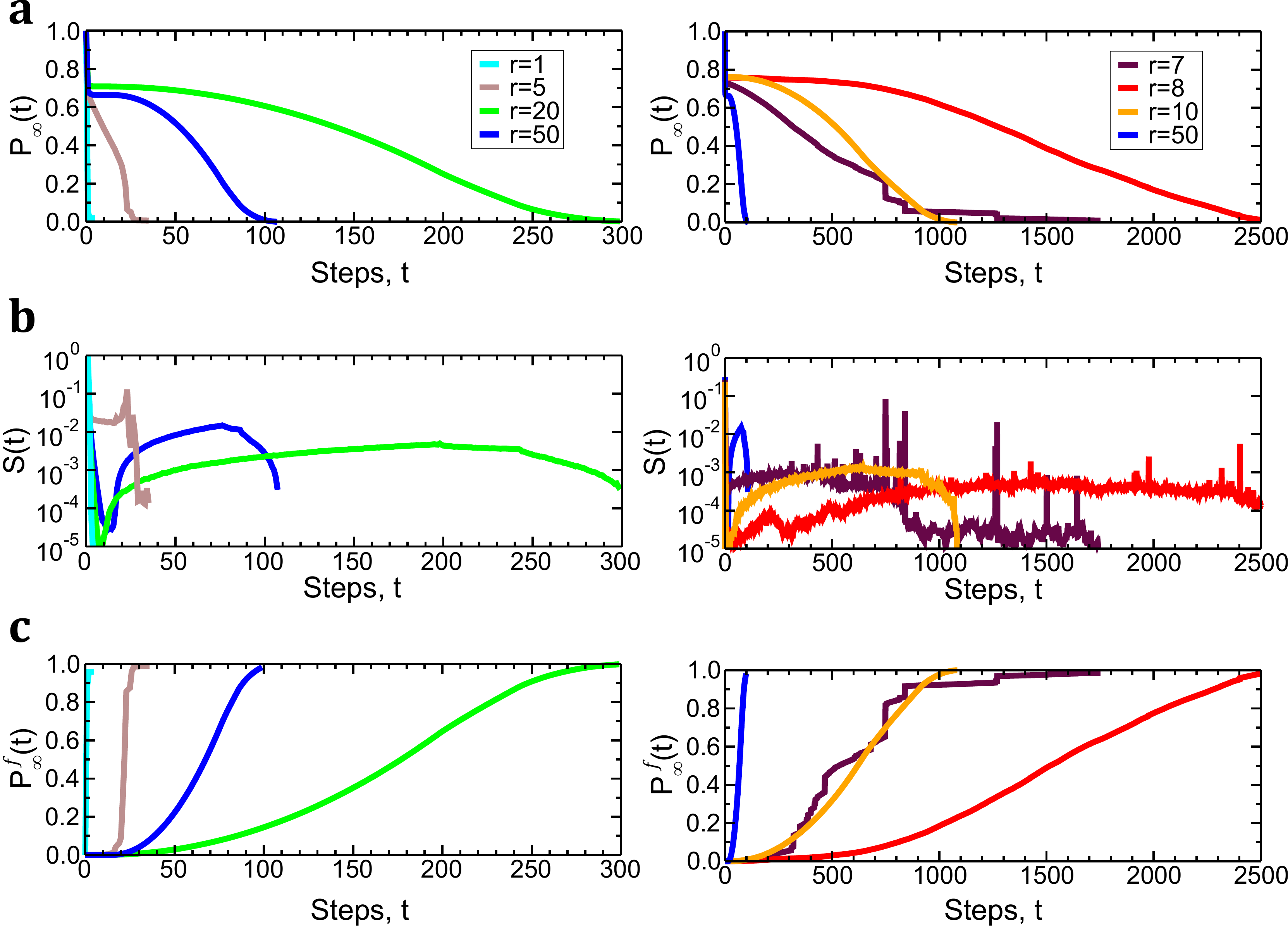}
    \caption{{\bf a,} Time evolution of the GCC, for different values of $r$. {\bf b,} The number of failed nodes per time step, $t$, {\bf c,} The size of the failed-GCC, as a function of time steps, $t$. All results are for $L=5000$.}
    \label{fig:S1}
\end{figure}

In Fig.~\ref{fig:S1}, we present the time evolution of the largest cluster size, $P_\infty(t)$, the fraction of failed nodes per step $S(t)$, and the size of the failed largest cluster, $P^f_\infty(t)$, in the range $r=1$ to $r=20$. These are the same quantities shown in Fig.~\ref{fig:2} for larger $r$. We reproduce the case of $r=50$ from Fig.~\ref{fig:2} in all plots, for comparison purposes.

\textbf{Regime (I)}. For small $r$ values ($r < r_c$), the failure process shown in Fig.~\ref{fig:S1} (e.g., $r=5$) is characteristic of a second-order percolation transition. The system fails at a high, almost constant rate with time, without any plateau behavior in the GCC size. The key distinction from a typical percolation transition is that, here, damage arises spontaneously from the system dynamics, and not by varying the external control parameter, i.e. removing an increasingly larger fraction of nodes. The similarity between the static percolation transition and the transition led by internal dynamics for small $r$ leads to the conclusion that in this regime this dynamic process is also removing nodes in a uniformly distributed manner. This behavior can be understood because the short dependency range always removes nodes in the vicinity of prior damage, so that all failures remain close in space. 

\begin{figure}[p!]
    \centering
    \includegraphics[height=\textheight]{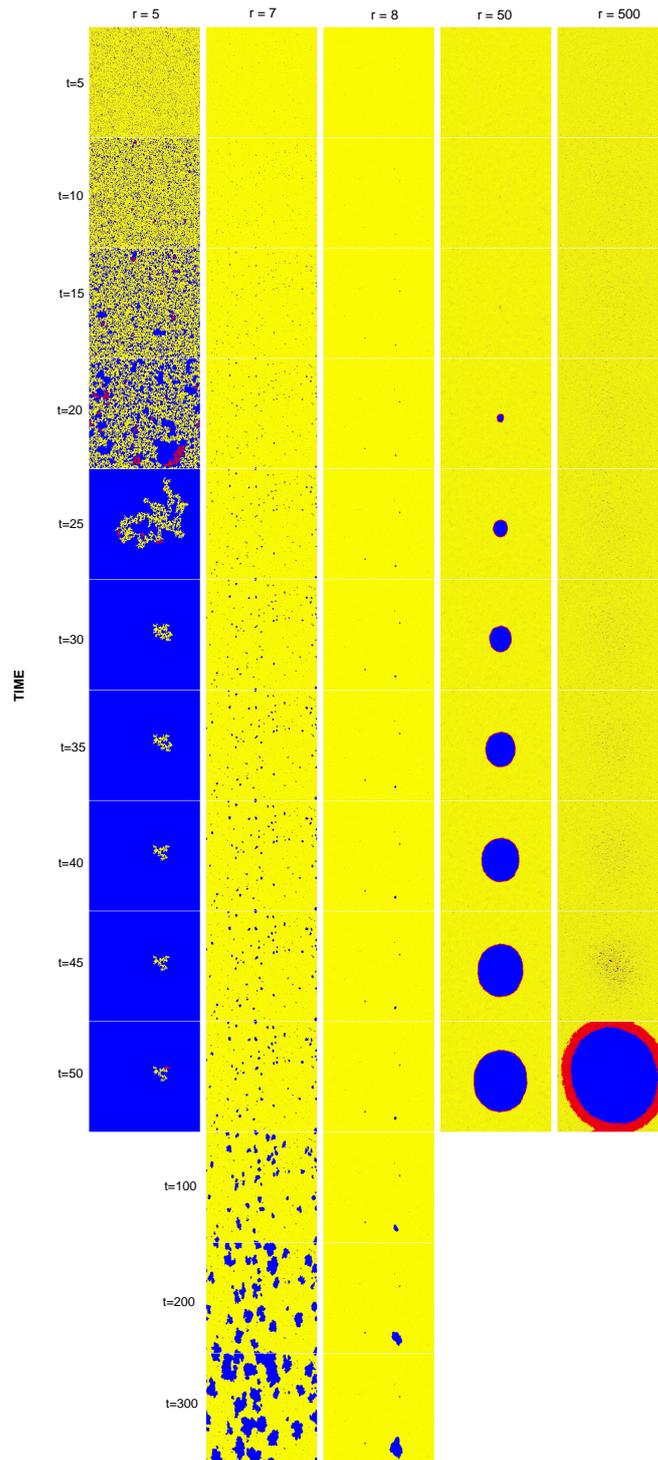}
    \caption{
    Snapshots of the time evolution of the collapse at $p_c$. Each column represents a different $r$ value and time increases from top to bottom. The yellow nodes represent the GCC, the red nodes are those that failed at the current step, and blue nodes represent all the failed nodes up to the given step. We use periodic boundary conditions and we shift the images so that the nucleus appears in the center of the system.}
    \label{fig:S2}
\end{figure}

This is also evident in Fig.~\ref{fig:S2} for $r=5$, where the collapse process with time looks very similar to the corresponding collapse of static percolation as we increase $p$. At early steps, nodes are removed randomly and as the damage increases with time, the GCC has clear fractal characteristics.

\textbf{Regime (II)}. As we increase $r$ towards the critical value $r_c=8$, and just before reaching it, i.e. when $r=7$, the value of $p_c$ increases significantly and the process has a much longer duration. As the dependency range is now larger, damage can spread to relatively further distances. This results in  multiple pockets of failure around the initially removed nodes. Since $p_c$ is now large, there is a smaller initial damage and the process develops slowly. The extended, but still limited, dependency range creates these pockets of damage uniformly in the system. This can also be understood by the GCC and the failed-GCC evolution in Fig.~\ref{fig:S1}. The jumps in the failed-GCC indicate that large pockets of damage coallesce and thus significantly decrease the GCC size. This mechanism is quite similar with smaller $r$ values, as the fraction of nodes removed per step is roughly constant, but at a microscopic level the collapse is due to the emergence of multiple nuclei which develop concurrently throughout the system (Fig.~\ref{fig:S2}).

\textbf{Regime (III)}. At the critical value, $r_c=8$, there are no longer multiple nuclei, but only one. Random fluctuations combined with the longer distances, which cause larger overlaps of damage, create an area of significantly lower density than in the rest of the system. The density remains relatively high throughout the lattice and the failure process would terminate at this high value of $p_c$. However, the early spontaneous emergence of the nucleus creates a damaged front which propagates through layers of width $r$. This process results in a smooth evolution of the GCC with time, where at early times the branching process creates a plateau of limited damage per step. The formation of the nucleus. however, increases the rate of failure as it increases in size through nucleation propagation. This can also be seen in Fig.~\ref{fig:S2} for $r=8$. It is clear, therefore, how the emergence of a nucleus leads to a first-order transition with $p$. As we lower $p$, the density decreases slowly and linearly with $p$. At a given point, $p_c$ which is still far from the normal percolation threshold, the nucleus propagation causes the system to collapse completely, causing the singularity that we observe for the GCC at $p_c$.

As $r$ increases further (e.g. $r=50$ in Fig.~\ref{fig:S1}, the basic mechanism of an initial branching process followed by nucleation propagation remains the same. Since the damage front is proportional to $r$, the propagation occurs faster, as described in the main text, and the nucleus size increases at a faster rate for larger $r$ (compare, for example, $r=8$ vs $r=50$ in Fig.~\ref{fig:S2}.

\textbf{Regime (IV)}. When $r$ increases significantly and approaches the system size, $L$, the nucleation mechanism emerges late in the process and lasts only a few steps. Any damage can now immediately cause damage almost anywhere in the entire system, so that local fluctuations are difficult to emerge. Therefore, the branching mechanism dominates the process the node density decays quite uniformly throughout the lattice. As a result, the reduction of the GCC size is gradual and follows the path of a second-order transition (similar to the behavior of the GCC size in static percolation as we approach the critical threshold from above). However, at the critical threshold $p_c$ the system density becomes sufficiently low for a nucleus to emerge (in a similar way as described for smaller $r$), but now its characteristic radius $r\sim L$ covers almost the entire system. This means that the nucleation event leads to the system collapse within a few steps, as there is no space for propagation. The initially continuous decay of $P\_infty(p)$ with $p$, which is followed by an abrupt collapse as soon as a nucleus appears, indicates that this is a typical model for a mixed-order PT.

Figure~\ref{fig:S2} summarizes and compares the difference between these four regimes, as the system collapses at the threshold $p_c$ (the value of $p_c$ is different in each case).  (I) For $r =5$, the GCC decays as a typical percolation process where damage remains highly local and the removed clusters have fractal characteristics. (II) When $r=7$, the process starts similarly, but the increased dependency distance now gives each growing cluster a higher probability of bridging local gaps. This leads to multiple nuclei with relatively smooth edges to grow at similar rates, until they merge and percolate across the system. (III) For $r = 8$($=r_c$) and $r=50$ the dynamic changes completely. The branching process becomes increasingly non-local and uniform, until the spontaneous appearance of one nucleus (compared to multiple nuclei for $r<r_c$. The system collapses after the nucleus propagates across the entire lattice, where the width of the damage front is proportional to $r$. (IV) For very large $r$, such as $r=500$, the same process results in a nucleus of the size of the system. This is an extreme version of the previous regime where the spontaneous nucleus already covers most of the system and has no space to propagate.
tinct regimes.

To summarize, the first two regimes, (I) and (II), rely exclusively on a branching process and yield a smooth second-order transition. The main difference in the microscopic features of the two regimes is the shape of the growing damaged clusters. In the other regimes, (III) and (IV), there is a long branching process which is followed by the spontaneous emergence of a nucleus that propagates. The two regimes differ in the size of this nucleus, which for large $r$ is of the order of the system size and has very little space to propagate. 

\subsection*{Local density}

As described in the previous section, the different failure regimes are controlled by how the local node density behaves and how uniform this density remains throughout the system during the damage propagation. To verify this, we numerically investigate the varying evolutionary patterns of the local density as we change $r$. Local density is here defined as the fraction of surviving nodes within a small network area compared to the initial number of nodes in that area of the original network.

\begin{figure}[tbh]
    \centering
    \includegraphics[width=\columnwidth]{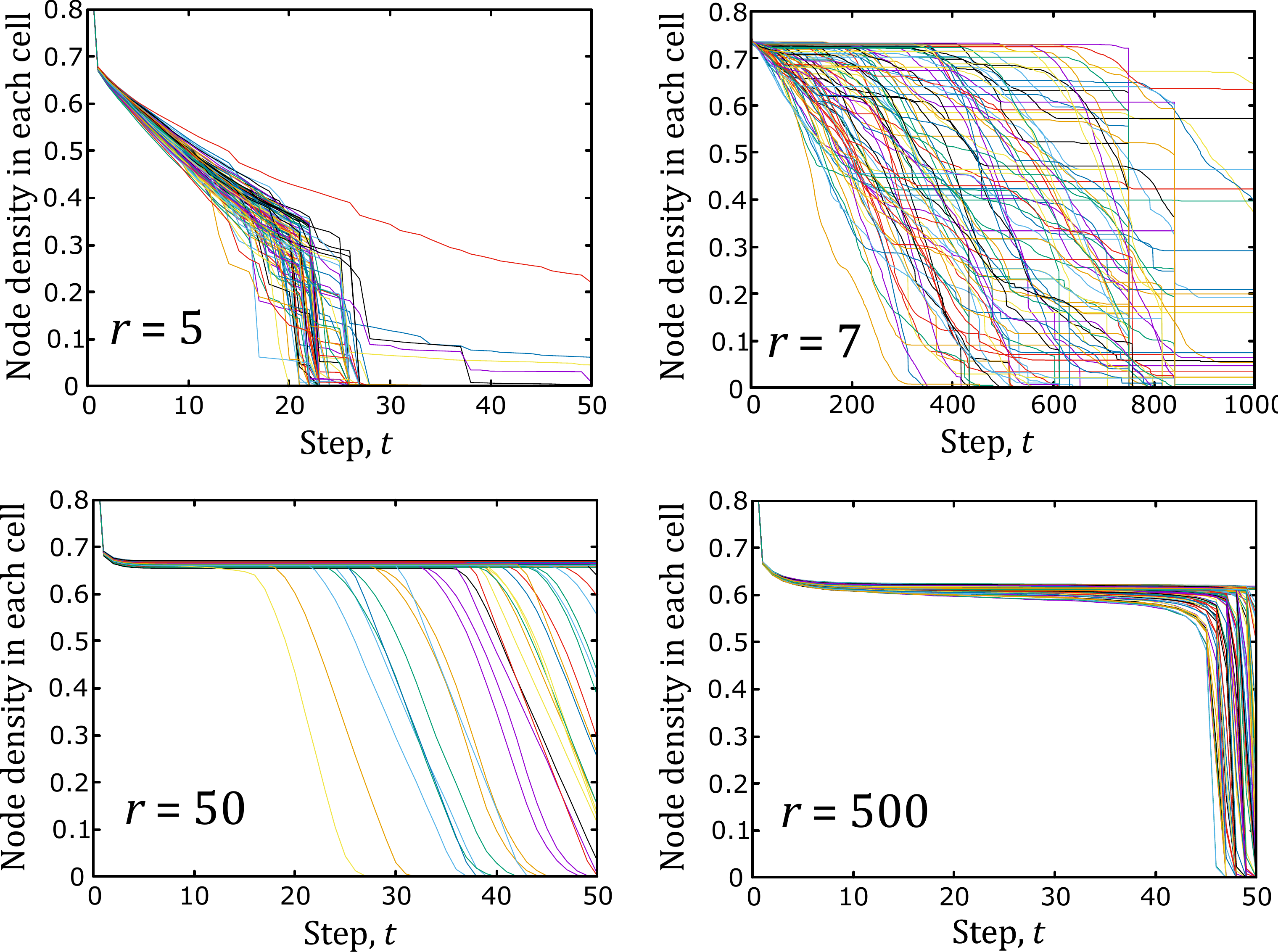}
    \caption{Time evolution of the local density in each of the 100 cells in a system of $L=5000$. The value of $r$ in each case is shown in the plot.}
    \label{fig:S3}
\end{figure}

We quantify the local density evolution by dividing the $L=5000$ lattice into a $10 \times 10$ square grid, creating 100 local non-overlapping cells. Each cell includes $500 \times 500$ sites, totaling $250,000$ nodes. The local density within each of these 100 cells is tracked over time by measuring the survival fraction of its nodes at each step. The resulting time series for the 100 local densities are shown for different values of $r$ in Fig.~\ref{fig:S3}.

The patterns of local density evolution for each $r$ value are distinct, reflecting the different microscopic failure mechanisms. For $r=5$, the local density decays uniformly across the system, independent of cell location, with the exception of the largest surviving cluster. In contrast, at $r=7$, the system shows significant heterogeneity: density drops sharply to zero in some cells while others remain high, a behavior which is consistent with simultaneous damage growth in multiple areas (notice also the different timescale). For $r=50$, the initial uniformity is maintained until the density in one cell decays rapidly, which marks the nucleation event. Following this, the density drops in other cells sequentially, consisten with nucleation propagation. Finally, at $r=500$, the local density remains uniformly high across all cells until the entire system undergoes a rapid collapse within a few time steps.

\begin{figure}[tbh]
    \centering
    \includegraphics[width=\columnwidth]{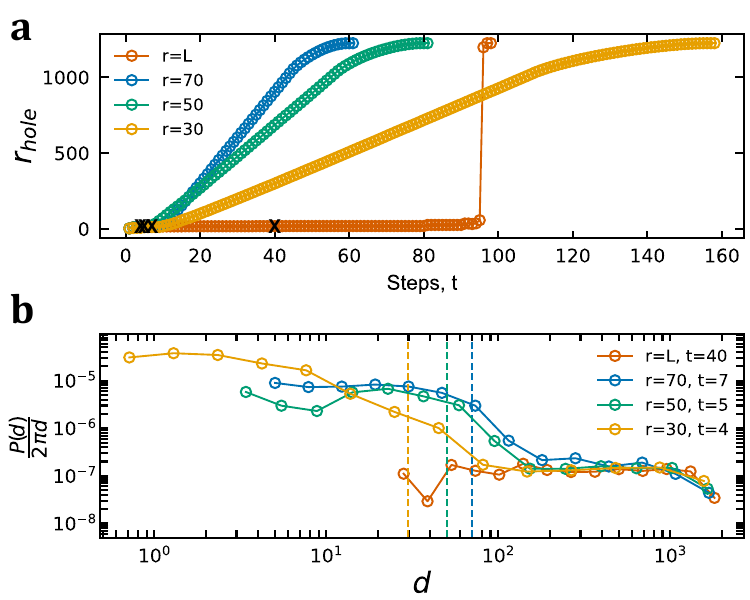}
    \caption{Localized branching process at finite-range
dependency ($r=30, 50, 70$) versus the global branching process for infinite-range dependency ($r=L$). {\bf a,} Evolution of the maximum
hole radius, $r_{\rm hole}$, during the critical cascading process. The branching mechanism prevails when $r_{\rm hole}$ remains zero, while the increase in the hole radius indicates nucleation propagation. The crosses indicate the step where we calculated the distributions in panel (b). {\bf b,} Rescaled distribution of distances from the hole center to failing nodes at specific time steps during the branching process, where $P(d)$
is scaled by $2\pi d$. The system size is L = 1000.}
    \label{fig:S4}
\end{figure}

\subsection*{Distribution of failures during the branching process}

To further illustrate the localized nature of the branching
process in finite $r$ cases, we analyze the rescaled distribution
of distances from a failing node to the center of the
developing nucleus. In Fig.~\ref{fig:S4}a, we can see the evolution of the hole radius at different $r$ values. As expected, the hole grows faster as we increase $r$ due to the wider damage front, except for the case of $r=L$ where there is no nucleus until the later stage of the process. Here, we want to focus on the branching process, i.e. before any significant hole emerges, so we select time steps where $r_{\rm hole}$ is still zero.

We then calculate the distance distribution of failed nodes from the center of the hole, at the selected time step. Notice that the hole has not yet developed at these early steps, but we use our knowledge of the hole location at later steps, to test whether nodes fail uniformly or not in the vicinity of the eventual hole. The distribution of failures for finite $r$ in Fig.~\ref{fig:S4}b indicates that there is a significantly higher probability of nodes failing within a distance $r$ from the hole center, rather than in the rest of the system. This concentration suggests
that node failures during the branching process are localized
near the area where the hole will form in subsequent
steps, setting the stage for nucleation through a localized
collapse.  In contrast, there is no spatial preference when $r=L$ and the probability distribution of failure is largely independent of the distance from the hole center. The homogeneous distribution of failing nodes across the system, reflects a global branching process. These observations
clearly demonstrate that the localized branching process
initiates the nucleation process, linking finite and infinite
dependency-range scenarios through a common underlying
mechanism.

\subsection*{Branching factor}

Fig.~\ref{fig:4}a shows that for $r=L$, the value of $\eta$ is a reliable measure of the system collapse and at criticality  the branching factor assumes a value of $\langle\eta\rangle=1$. On the other hand, in finite-range dependencies (Fig.~\ref{fig:4}b) the critical branching factor is less than 1, which means that localized branching processes within a confined area can still trigger a nucleation process even though the global branching factor would indicate that the cascade stops.

This behavior demonstrates the critical role of the branching process in driving cascading failures. Here, we examine the evolution of the branching factor $\eta(t)$ during this cascading process with time, and we clarify how $\eta$ controls the system collapse.

\begin{figure}[tbh]
    \centering
    \includegraphics[width=\columnwidth]{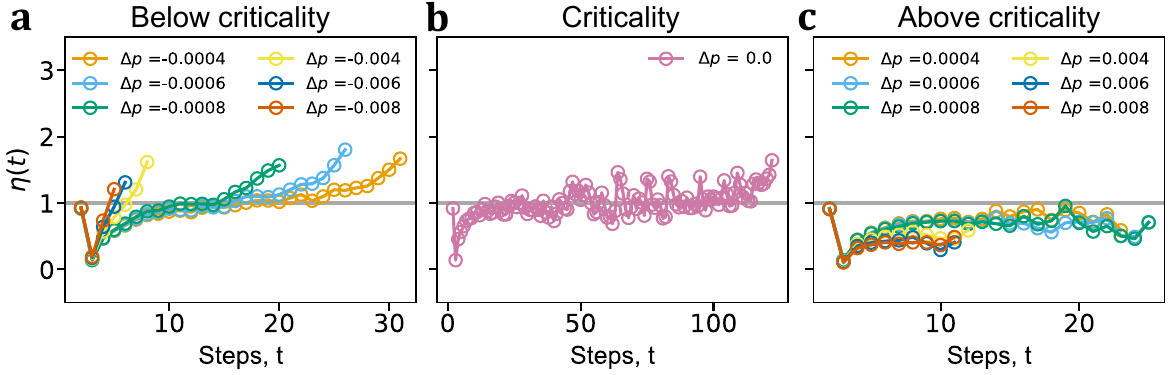}
    \caption{Evolution of the branching factor for $r=L$. {\bf a,} The Branching factor $\eta(t)$ as a function of time $t$ for various values of $p$ below $p_c$, where $\Delta p=p-p_c$. {\bf b,} The values of $\eta(t)$ at the critical point $p_c$. {\bf c,} The evolution of the branching factor above criticality. The lattice size is $L=1000$.}
    \label{fig:S5}
\end{figure}

We perform numerical simulations to investigate the evolution of the branching factor when $r=L$ at subcritical, critical, and supercritical points. At the critical point $p_c$, $\eta$ is practically constant and equal to 1 (Fig.~\ref{fig:S5}b). This indicates that the system has entered a critical branching process, where each failure triggers exactly one subsequent failure. In this stage, the system is balanced on the verge of collapse: small perturbations can propagate widely without causing global damage. When $p<p_c$, the branching factor
is initially below 1, but stabilizes around 1 for some time, before eventually increasing to even larger values (Fig.~\ref{fig:S5}a). This behavior describes an expanding cascade where each failure step removes more nodes than the previous one. This self-reinforcing process pushes the system
towards an inevitable collapse. When $\eta(t)$ surpasses
1, the system loses control over the failure propagation,
leading to rapid and cascading failures
throughout the network.

Conversely, for $p>p_c$, the branching factor stabilizes to a  value below 1 (see Fig.~\ref{fig:S5}c). While failures persist,
the cascading process weakens over time and eventually
halts. The system absorbs these disturbances without
undergoing total collapse, maintaining ita overall stability.
As $p$ increases from below $p_c$, the
evolution of $\eta(t)$ reflects a gradual progression from system collapse to stability.
These observations provide key insights into the mechanisms
driving system collapse in networks with infinite
dependency ranges.

\end{document}